\documentstyle[aps,twocolumn,epsfig,amsfonts]{revtex}

\draft 

\begin{document}

\title{Quantized vortices and collective oscillations\\ 
of a trapped Bose condensed gas}
 
\author{Francesca Zambelli and Sandro  Stringari}
\address{Dipartimento di Fisica, Universit\`{a} di Trento,}
\address{and Istituto Nazionale per la Fisica della Materia,}
\address{I-38050 Povo, Italy}
  
\date{\today}
 
\maketitle
 
\begin{abstract}
\noindent
Using a sum rule approach we calculate the frequency shifts of the
quadrupole oscillations of a harmonically trapped Bose gas 
due to the presence of a quantized vortex. 
Analytic results  are obtained for positive scattering lengths and large $N$
where the shift relative to excitations of opposite angular momentum is found 
to be proportional to the quantum circulation of the vortex and to decrease as 
$N^{-2/5}$. Results are  also given for smaller values of $N$ covering the 
transition between the ideal gas and the Thomas-Fermi limit.
For negative scattering lengths we predict a macroscopic instability of the
vortex. The splitting of the collective frequencies  in toroidal
configurations is also discussed.

\end{abstract}
 
\pacs{PACS numbers: 03.75.Fi, 05.30.Jp, 32.80.Pj, 67.40.Db}

\narrowtext

\noindent
After the experimental realization of Bose-Einstein condensation in
dilute atomic gases, the study of the collecive
excitations of these unique inhomogeneus quantum  systems has been
the object of  both  experimental \cite{Jin,Mewes,Jin2,Dan}
and theoretical \cite{Singh,Stringari} work (for an  update list of references 
see, for example, \cite{georgia}). These oscillations are
characterized by proper quantum numbers, reflecting the symmetry of the
confining potential. 
In an axially symmetric trap the third component of angular momentum is a 
natural quantum number and if the system is in a time reversal invariant 
configuration, elementary excitations carrying opposite angular momentum are
degenerate. This degeneracy is in general removed if time reversal symmetry 
is broken. 
The purpose of this work is to describe the frequency shifts produced
by the presence of a quantized vortex.
In view of the important role played by vortices in  understanding the 
mechanisms of superfluidity, the possibility of their spectroscopic 
diagnostics  is highly interesting since the measurements of collective 
frequencies can be  carried out with high precision in these systems.

The occurrence of splitting in the presence of a vortex can
be simply understood by noting that the average velocity
flow associated with the collective oscillation can be either 
parallel or  opposite to the vortex flow, depending on the 
sign of the angular momentum carried by  the excitation. This
produces a shift of the collective frequency of order
$\delta\omega/\omega\sim v/c$
where $v \sim 1/R$ is the velocity of the vortex flow and $c$
is the sound velocity. In a trapped Bose gas $c$ increases linearly
with the radius $R$ of the condensate, while $\omega$ is practically
independent of $R$, so one expects relative shifts of the order of $1/R^2$. 
These effects are larger than the typical corrections to the Thomas-Fermi 
limit due to finite size effects which, in the
absence of the vortex, behave like $logR/R^4$ \cite{Fetter&Feder}. 
The problem of the frequency shift produced by a quantized vortex 
has been already the object of theoretical work using
semiclassical approaches based on a large $N$ expansion  \cite{Sinha}, 
as well as by  full numerical solution of the
linearized equations of motion \cite{Burnett}.

In this  letter we develop a sum rule  approach \cite{Lipparini} which 
is expected to provide exact results for the splitting of the
excitation frequencies in large systems. The method can be also applied 
to calculate the frequency shifts for small values of $N$ as well as for
negative scattering lengths. 
Let us introduce the strength distribution function
\begin{equation}
S_{\pm}(E)=\sum_n |\langle n\;|F_{\pm}|\;0\rangle|^2 \delta
\big(E-\hbar\omega_{n0}\big)
\label{strength}
\end{equation}
relative to the operators
$F_{\pm}=\sum_{k=1}^N f_{\pm}({\bf r}_k)$
carrying opposite angular momenta.
In equation (\ref{strength}) $\hbar\omega_{n0}=(E_n -E_0)$ are the
excitation energies relative to the  eigenstates $|n\rangle$ of the
Hamiltonian
\begin{equation}
H = \sum_i \left({1\over 2M} p^2_i + V_{\rm ext}({\bf r}_i)\right)  + 
g\sum_{i < j}\delta({\bf r}_i-{\bf r}_j)\;,
\label{H}
\end{equation} 
which describes $N$ interacting bosons confined  by an external potential.
The potential 
$V_{\rm ext}({\bf r})\!=\!M(\omega_{\perp }^2r_{\perp }^2+\omega_z^2z^2)/2$, 
with $r_{\perp }^2\!=\!x^2+y^2$, is assumed to be axially symmetric, 
and the interatomic force is a contact 2-body interaction whose coupling 
constant $g=4\pi\hbar^2a/M$ is fixed by the $s$-wave scattering length $a$.

In the following we will focus on the collective oscillations 
of low multipolarity which are easily excited in experiments by suitable 
modulation of the harmonic trap. 
For the quadrupole case  we will consider the modes excited by the operators 
\begin{equation}
f_{\pm}=(x\pm iy)^2
\label{qpolov}
\end{equation}
and
\begin{equation}
f_{\pm}=(x\pm iy)z\;,
\label{qpolov1}
\end{equation}
carrying angular momentum $m\!=\!\pm 2$  and $m\!=\!\pm 1$ respectively 
(here and in the following we will identify  $m$ with the third component 
of angular  momentum of the elementary excitation).
Only excitations with $m\ne 0$ are relevant for the present discussion.

In the absence of vortices the ground state has zero angular
momentum and, for large $N$ and positive scattering lengths, the collective
states excited by the operators (\ref{qpolov}) and (\ref{qpolov1}) are 
well described by  hydrodynamic theory of superfluids. 
This yields \cite{Stringari} the result 
$\omega_{\pm}\!\!=\!\sqrt{2}\omega_{\perp}$ 
and $\omega_{\pm}\!\!=\!\sqrt{\omega_{\perp}^2+\omega_z^2}$ 
for the  $m\!=\!\pm 2$  and $m\!=\!\pm 1$ frequencies.  
Notice that these results differ from the ideal gas predictions,
$\omega_{\pm}\!\!=\!2\omega_{\perp}$ and 
$\omega_{\pm}\!\!=\!\omega_{\perp}+\omega_z$,
as a consequence of interaction effects which suppress
the contribution of the kinetic energy pressure term in the equations of 
motion. Differently from the quadrupole excitations, the frequencies of the
dipole modes excited by $f_{\pm}\!=\!x\pm iy$
are instead unaffected by two body interactions and are given by 
$\omega_{\pm}\!\!=\!\omega_{\perp}$. This behavior is the consequence 
of the translational invariance of the interatomic force
which cannot affect the motion of the center of mass,   
even in the presence of a vortex.

The moments 
\begin{eqnarray}
m_p^{\pm}&=&\int_0^{\infty}\!\!\!dE\big(S_+(E)\pm S_-(E)\big)E^p
\label{mp}
\end{eqnarray} 
of the strength distribution (\ref{strength}) can be calculated using
closure relations.
For the lowest moments we find the results:
\begin{eqnarray}
m_0^-&=&\langle [F_-,F_+]\rangle =0\label{m0c}\\
m_1^+&=&\langle \big[F_-,[H,F_+]\big]\rangle =\frac{N\hbar^2}{M}
\langle \;|\mbox{\boldmath $\nabla$}f_+|^2\rangle
\label{m1c}\\
m_2^-&=&\langle \big[[F_-,H],[H,F_+]\big]\rangle =
N\langle [j_-,j_+]\rangle\;,
\label{m2c}
\end{eqnarray}
where the average $\langle\;\rangle$ is taken on the state $|0\rangle$
which may or  may not contain a vortex and we have used the property 
$F_+^{\dagger}=F_-$. 

The first commutator (\ref{m0c}) vanishes  because the operators 
$F_+$ and  $F_-$,  depend only on the  spatial coordinates.
The double commutator (\ref{m1c}) is the analog of the 
f-sum rule \cite{Pines} and gets  contribution only from the kinetic
energy term since both the external potential and the two-body
interaction commute with $F_{\pm}$.
Finally the current operators $j_{\pm}=[p^2/2M,f_{\pm}]$ entering the third 
sum rule are defined by
\begin{equation}
j_{\pm}=\frac{\hbar}{2Mi}\mbox{\boldmath $\nabla$}f_{\pm}
({\bf r})\cdot{\bf p}+ h.c.\;,
\label{J+-}
\end{equation}
where ${\bf p}$ is the usual momentum operator.
Evaluation of the sum rules $m_1^+$ and $m_2^-$ is straightforward
in the case of the quadrupole operators (\ref{qpolov}) and (\ref{qpolov1}). 
For $m\!=\!\pm 2$ we find the result
\begin{eqnarray}
m_1^+&=&\frac{8\hbar^2}{M}N\langle r^2_{\perp}\rangle
\label{m1qpolo}\\
m_2^-&=&\frac{16\hbar^3}{M^2}N\langle l_z\rangle =
\frac{16\hbar^4}{M^2}N\kappa\;,
\label{m2qpolo}
\end{eqnarray}
where $l_z$ is the $z$th component of the angular momentum 
operator and $\kappa$ is the quantum of circulation of the vortex
($\kappa =\pm\;1,\pm\;2,...$).
Of course in the absence of vortices the sum rule $m_2^-$ vanishes. 
Notice that results (\ref{m1qpolo}-\ref{m2qpolo}) are formally independent of 
the choice of the external potential as well as of the two-body interaction. 

The  results for the moments $m_0^-$, $m_1^+$ and $m_2^-$
can be used to calculate the shift of the collective frequencies 
when the number of atoms in the trap is large and $a$ is positive. 
In fact in this limit, where the behavior of the system is properly described 
by hydrodynamic theory of superfluids, one expects that the moments calculated
above will be exhausted by two modes with frequency $\omega_{\pm}$ excited,
respectively, by the operators $F_{\pm}$:
\begin{equation}
S_{\pm}(E)=\sigma^{\pm}\delta (E-\hbar\omega_{\pm})\;,
\label{sigma}
\end{equation}
where $\sigma^{\pm}$ are the corresponding strengths.
Assumption (\ref{sigma}) is equivalent to a Bijl-Feynman ansatz \cite{Feynman}
often used to describe the collective excitations in interacting many-body
systems. In the case of superfluid helium it provides an exact description of
the excitation spectrum in the phonon regime (for a recent discussion on sum
rules and collective excitations in Bose superfluids see, for example,
\cite{Stringari93}).

Let us discuss the consequence of the vanishing of the $m_0^-$ moment 
(\ref{m0c}). With assumption (\ref{sigma}) for the strength distribution one
immediately finds the result $\sigma^+\!\!=\!\sigma^-$ and the
splitting between the two frequencies can be directly written as
\begin{equation}
\hbar (\omega_+-\omega_-)= m_2^-/m_1^+\;.
\label{21}
\end{equation}
Use of (\ref{m1qpolo}-\ref{m2qpolo}) then yields the relevant result
\begin{equation}
\omega_+\!\!-\omega_-\!=\frac{2}{M}
\frac{\langle l_z\rangle}
{\langle r^2_{\perp} \rangle}=\frac{7\omega_{\perp}\kappa }{
\lambda^{2/5}}\bigg(\!15\frac{Na}{a_{\perp}}\bigg)^{-2/5}
\label{hdshift}
\end{equation}
for the $m\!=\!\pm 2$ modes, where 
$a_{\perp}\!\!=\!\!\sqrt{\hbar /M\omega_{\perp}}$
is the oscillator length in the radial direction, while 
$\lambda =\omega_z/\omega_{\perp}$ characterizes the deformation of the 
harmonic trap.
The same calculation can be carried out for the modes excited by the 
$m\!=\!\pm 1$ quadrupole operators (\ref{qpolov1}). 
In this case the result is
\begin{equation}
\omega_+\!\!-\omega_-\!=\frac{2}{M}\frac{\langle l_z\rangle}
{\langle r^2_{\perp}+2z^2\rangle }
=\frac{7\omega_{\perp}\kappa\lambda^{8/5}}{1+\lambda^2}
\bigg(\!15\frac{Na}{a_{\perp}}\bigg)^{-2/5}\;.
\label{hdm=1shift}
\end{equation}
In the last equalities of (\ref{hdshift}-\ref{hdm=1shift})
we have used the Thomas-Fermi approximation \cite{Baym} to evaluate the square
radii of the condensate.
Notice that for spherical trapping the splitting between the $m\!=\!\pm 2$ 
frequencies is twice the splitting between the $m\!=\!\pm 1$ modes.
For large $N$ the shifts  become smaller and smaller showing that that in this
limit the  effects associated with the current of the  vortex are small 
corrections to the collective flow of the oscillation.
Neverthless the splittings can be sizable.
For example using a spherical configuration with $a/a_{\perp}\!=\!10^{-3}$
and $N=10^6$ and taking one quantum of circulation ($\kappa\!=1$) 
one finds that the relative shift $(\omega_+-\omega_-)/\omega$ of the
$m\!=\!\pm 2$ states is about $10\%$, having used the large $N$ result
$\sqrt{2}\omega_{\perp}$ for the average frequency $\omega$.
This shift is much larger than  the typical experimental uncertainties in
the measurements of the collective frequencies \cite{Jin,Mewes,Jin2,Dan}.

It is worth pointing out that the frequency shift due to the vortex is
exhibited by the quadrupole excitations, but not by the dipole modes excited
by the operators $f_{\pm}\!\!=\!\! x\pm iy$. In fact in the dipole case the 
operators $j_{\pm} =\hbar(p_x\pm iy)/Mi$ commute and the sum rule $m_2^-$ 
identically vanishes. As a consequence one finds $\omega_+\!\!=\omega_-$ as 
expected from general arguments.
Notice however that in addition to the above modes excited by the center
of mass operators, another dipole mode, localized near the core of the vortex,
has been predicted \cite{Burnett} to occur with frequency different from
$\omega_{\perp}$. It has been recently suggested \cite{Rokhsar} that this mode
could play an important role in driving the instability of the vortex.

The above results for the shift of the quadrupole frequencies
hold for large $N$, where the assumption that the operators 
$f_+$ and $f_-$ excite a single mode is justified.  
When the adimensional parameter $Na/a_{\perp}$ becomes small, this assumption
is no longer valid. In particular, in the limit of a non interacting gas, 
a vortex with quantum circulation $\kappa\!\!=\!+1$ corresponds to putting all
the atoms in the $1\;p$ state ($l_z\!=\!+1$) of the harmonic oscillator 
hamiltonian and the $m\!=\!-2$
operator $f_-$ can give rise to $\delta\omega\!=\!2\omega_0$ 
as well as to $\delta\omega\!=\!0\omega_0$ transitions. 
Viceversa the operator $f_+$ gives rise only to 
$\delta\omega\!=\!2\omega_0$ transitions.
The corresponding strengths are, respectively, $\sigma^-_{\rm up}=
2a_{\perp }^4N$, $\sigma^-_{\rm down}=4a_{\perp }^4N$ and
$\sigma^+=6a_{\perp }^4N$.

In order to study the transition from the noninteracting to the large $N$
regime we have to remove the single-mode assumption (\ref{sigma})
and hence we need the knowledge of  additional moments 
of the distribution function. The moments $m_3^+$, $m_4^-$ and $m_5^+$ 
can be easily calculated  for the quadrupole
operators. For example, for $m=\pm\;2$,  we find 
\begin{eqnarray}
m_3^+&=&\frac{16\hbar^4\omega_{\perp}^2}{M}N\langle r^2_{\perp}\rangle
\bigg[1+\frac{E_{\rm kin_{\perp}}}{E_{\rm ho_{\perp}}}\bigg]
\label{m3r}\\
m_4^-&=&\frac{64\hbar^5\omega_{\perp}^2}{M^2}N\langle l_z\rangle =
\frac{64\hbar^6\omega_{\perp}^2}{M^2}N\kappa
\label{m4r}\\
m_5^+&=&\frac{32\hbar^6\omega_{\perp}^4}{M}N\langle r^2_{\perp}\rangle
\bigg[1+3\frac{E_{\rm kin_{\perp}}}{E_{\rm ho_{\perp}}}+
\frac{\tilde{V}_{\rm int}}{16E_{\rm ho_{\perp}}}\bigg]\;,
\label{m5r}
\end{eqnarray}
where $E_{\rm kin_{\perp}}$ and $E_{\rm ho_{\perp}}$ are the radial 
contributions to the kinetic and oscillator energies respectively, and 
$$\tilde{V}_{\rm int}=g\;a_{\perp}^4\!\int\!d{\bf r}
\bigg[\frac{8\kappa^2n_0^2}{r^4_{\perp}}+ 
\nabla_{\perp}^2n_0\bigg(\frac{|\mbox{\boldmath $\nabla$}_{\perp}n_0|^2 }{n_0}
- \nabla^2_{\perp}n_0\bigg)\bigg]$$
is the contribution to $m_5^+$ arising from two-body interactions 
where $n_0$  is the density of the condensate and $\kappa$ is the quantum of
circulation of the vortex.
Result (\ref{m3r}) for  $m_3^+$ permits to obtain directly the asymptotic 
behavior  of the quadrupole frequency. 
In fact in the large $N$ limit the kinetic energy term in (\ref{m3r}) is
negligible and the ratio $(m_3^+/m_1^+)^{1/2}$ approaches the hydrodynamic
result $\sqrt{2}\omega_{\perp}$. 
Result (\ref{m4r}) for $m_4^-$ provides a crucial check of the validity of
(\ref{hdshift}).
In fact one can see that in the large $N$ limit the single-mode approximation
(\ref{sigma}), with the dispersion law    
\begin{equation}
\omega_{\pm}=\sqrt{2}\omega_{\perp} \pm 
\Delta\omega \;,
\label{omegapmdelta}
\end{equation}
where $\Delta\omega=\omega_+-\omega_-$ is given by (\ref{hdshift}), is
consistent with result (\ref{m4r}) for the sum rule $m_4^+$, up to effects
linear in $\Delta\omega$.
Differently from the other sum rules, $m_5^+$ depends explicitly on
the two body interaction. 
This contribution is very sensitive to the core region of the vortex and is
important to describe the crossover  from the noninteracting to the
Thomas-Fermi limit.

We have calculated numerically the sum rules $m_p^{\pm}$ with $p\!=\!1,..,5$ 
using the solution of the Gross-Pitaevskii equation for a vortex of 
quantum circulation $\kappa\!=\!+1$ \cite{Dalfovo}.
These sum rules are then used to evaluate the quadrupole
energies and strengths, by assuming that the operator $f_-$ excites two states
with different energy, in analogy with the behavior of the noninteracting
model.
The results for a spherical potential are reported in Figure 1.
where, for semplicity, we have plotted only the low frequency mode 
excited by $f_-$. 
The strength relative to the high frequency mode excited by $f_-$  in fact
vanishes rapidly when $N$ increases and hence this mode 
is not  physically relevant, except for small values of $N$. 
In the Figure we also report the average energy
$(m^+_3/m^+_1)^{1/2}$ calculated in the absence of vortices. This energy
turns out to be very close to the exact numerical solution of 
the linearized Gross-Pitaevskii equation (differences are always smaller than 
$0.5\%$). Notice that the frequency $\omega^-_{\rm down}$ corresponds to a 
rigorous upper bound to the lowest frequency of modes excited by $f_{\pm}$.

The frequencies $\omega_{\pm}$ start respectively from $2\omega_{\perp}$ and 
$0\omega_{\perp}$ and approach the value $\sqrt{2}\omega_{\perp}$
for large $N$. In the same limit the  strengths $\sigma^{\pm}$
tend to  the asymptotic value
$\sigma^+\!\!=\!\sigma^-\!\!=\!4\sqrt{2}\;a_{\perp}^4N
\lambda^{2/5}(\!15Na/a_{\perp})^{2/5}/7$.
The Figure  shows that for large $Na/a_{\perp}$ the asymptotic
dispersion relation (\ref{omegapmdelta}) well reproduces the behavior of the
two collective frequencies. 
Notice that for such values of $Na/a_{\perp}$ the strengths 
$\sigma^{\pm}$ practically coincide, confirming the validity of the single 
mode assumption (\ref{sigma}). 
The small asymmetry of the calculated frequencies with respect to the 
hydrodynamic value is a consequence of the fact that for the values
of $Na/a_{\perp}$ reported in the figure the excitation energy differs 
from $\sqrt2\omega_{\perp}$ even in the absence of vortices. 

We have also explored the case of negative scattering lengths. In this case we
find that the excitation energy of the lowest mode becames negative revealing
that the vortex is unstable against macroscopic fluctuations of the density.
The dispersion law for small value of the parameter $a$ is found to be
\begin{equation}
\omega^-_{\rm
down}=\omega_{\perp}\bigg(\frac{Na}{a_{\perp}}\bigg)\sqrt{
\frac{\lambda}{2\pi}}
\label{aneg}
\end{equation}
and changes sign with $a$.

It is finally interesting to discuss how the above results are modified by 
changing the geometry of the problem.
This is important because toroidal configurations are expected to suppress
the mechanisms of instability of the vortex \cite{Rokhsar}.
A useful way to pin a vortex might be  achieved with a thin laser beam
stabilizing the core of the vortex  along  the $z$-axis.
The thickness $d$ of the beam can be a few microns. This is larger than the
size of the core,  fixed by the coherence length $\xi=a_{\perp}^2/R$, 
but can be significantly smaller than the size of the condensate $R$.
In this case the structure of the core of the vortex will be significantly
modified by the pinning, but the macroscopic behaviour of the collective
excitations and in particular results (\ref{hdshift}-\ref{hdm=1shift}) for
the splitting, will be modified only in a minor way. 
An estimate of this effect can be obtained by calculating the change of
$\langle r^2_{\perp}\rangle$ due to the presence of the repulsive
potential generated by the laser beam. We expect small effects if $d \ll R$. 

A quite different behaviour is achieved by choosing a ring geometry. 
In the case of an ideal ring of radius $R$ the problem
is analytically soluble using Bogoliubov theory \cite{Rokhsar}. 
In this case the natural excitation operators have the form 
$f_{\pm}=exp(\pm i m\phi)$ where $\phi$ is the azimuthal angle and $\pm m$ is 
the angular momentum carried  by the excitation. The kinetic energy operator 
can be replaced by $(-\hbar^2/2MR^2)\partial^2/\partial\phi^2$ and the sum 
rules (\ref{m1c}) and (\ref{m2c}) become $m_1^+=\hbar^2m^2/MR^2$ and 
$m_2^-=2\hbar^3m^3\langle l_z\rangle /M^2R^4$.
Using the Bijl-Feynman ansatz (\ref{sigma}) one immediately finds 
$\omega_+-\omega_-=2\hbar m\kappa /MR^2$, where $\kappa$ is the quantum of 
circulation of the vortex, in agreement with the results 
recently discussed in \cite{Rokhsar}.
Notice that for large $R$ the lowest collective excitations in the ring 
geometry correspond to one-dimensional compression waves with dispersion 
\begin{equation}
\omega= \frac{c\;|m|}{R} \pm \frac{\hbar\;\kappa\;m}{MR^2}
\label{ring}
\end{equation}
where $c$ is the sound velocity. These frequencies, which are the analog of
(\ref{omegapmdelta}), coincide with the ones calculated for a system at rest
in a frame rotating with angular velocity $\omega =\hbar\kappa /MR^2$.

After completing this paper we received a preprint by A.A. Svidzinsky and 
A.L. Fetter (cond-mat/9803181), based on a hydrodynamic description of normal 
modes. Similarly to the results of \cite{Sinha}, this work predicts a frequency
shift also for the dipole mode. The origin of this discrepancy remains to be
understood.

Useful discussions with F. Dalfovo, A.L. Fetter, M. Gunn and N. Wilkin-Gunn are
acknowledged. This work was supported by the BEC advanced research project of
INFM.


\noindent
FIGURE CAPTION:

\noindent
{Frequencies (a) and strengths (b) relative to the $m= \pm 2$ 
quadrupole modes in the presence of a $\kappa =1$ vortex, as a function
of $Na/a_{\perp}$ for a spherical trap ($\omega_{\rm ho}=\omega_{\perp }$,
$a_{\rm ho}=a_{\perp }$, $\lambda =1$). 
The dotted lines correspond to the large $N$ behavior (\ref{omegapmdelta}).
The arrow indicates the Thomas-Fermi limit $\omega = \sqrt2 \omega_{\perp}$.
Dashed-dots correspond to the ratio $(m_3^+/m_1^+)^{1/2}$ without vortex.
Strengths are given in units of $a_{\perp}^4N$.} 


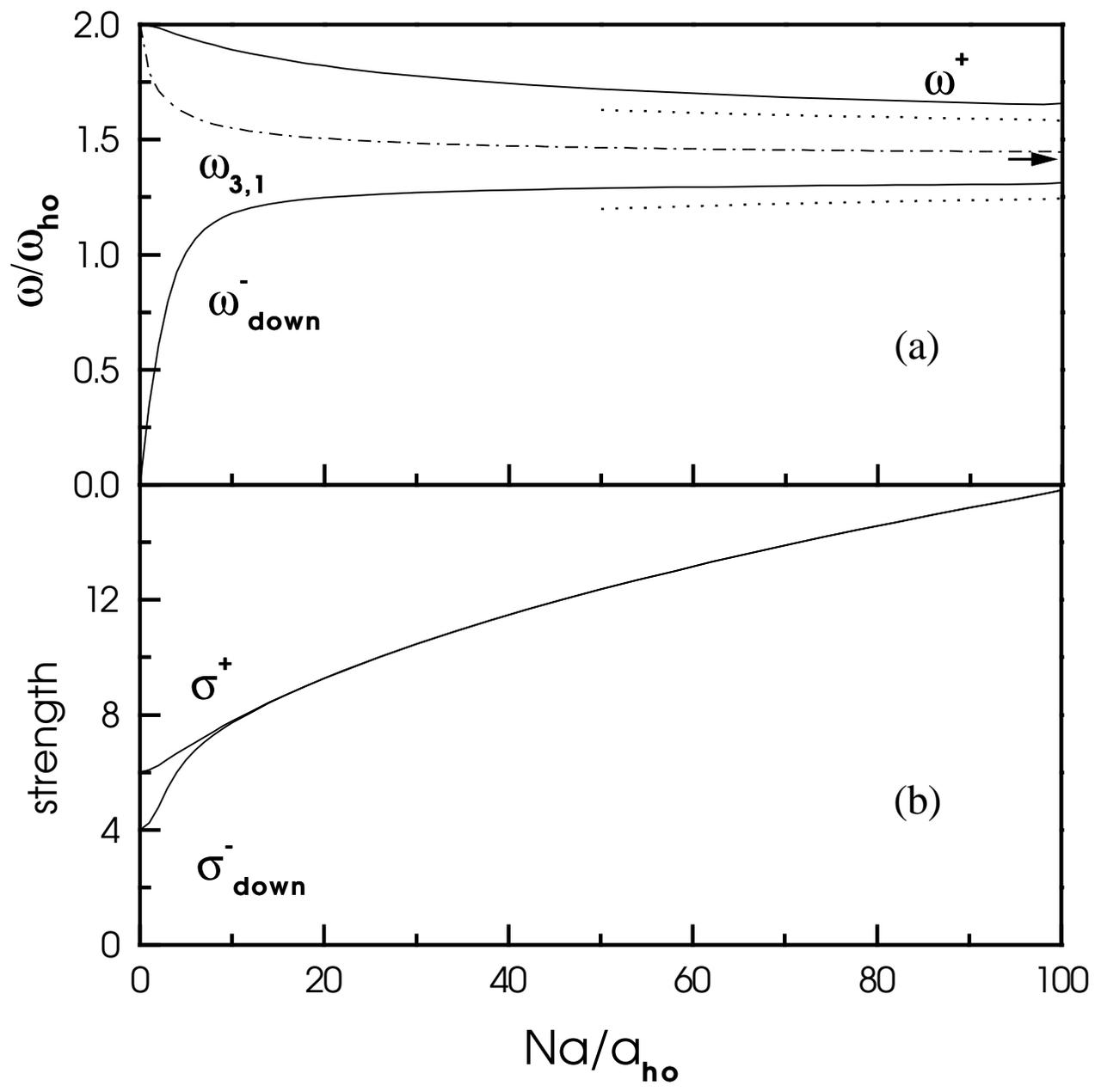
\end{document}